\documentclass[epj]{svjour}

\usepackage[utf8x]{inputenc}

\usepackage{graphics}
\usepackage{amsmath}
\usepackage{amsfonts}
\usepackage{epsfig,lscape,setspace,url, color, ulem}

\newcommand \ba  {\begin{array}}
\newcommand \ea  {\end{array}}
\newcommand \bi  {\begin{itemize}}
\newcommand \ei  {\end{itemize}}
\newcommand \ben  {\begin{enumerate}}
\newcommand \een  {\end{enumerate}}
\newcommand \be  {\begin{equation}}
\newcommand \bea {\begin{eqnarray}}
\newcommand \ee  {\end{equation}}
\newcommand \eea {\end{eqnarray}}

\begin{document}

\title{The Lehman Brothers Effect and Bankruptcy Cascades}

\author{Paweł Sieczka\inst{1,}\thanks{e-mail: psieczka@if.pw.edu.pl}, 
Didier Sornette\inst{2,}\thanks{e-mail: dsornette@ethz.ch}, 
and Janusz A. Hołyst\inst{1,}\thanks{e-mail: jholyst@if.pw.edu.pl} }

\institute{Faculty of Physics, Center of Excellence for Complex Systems Research,
 Warsaw University of Technology,\\ Koszykowa 75, PL-00-662 Warsaw, Poland 
\and ETH Zurich, Switzerland, and Swiss Finance Institute, Switzerland
\\
\\
Dedicated to Werner Ebeling on the occasion of his 75th birthday}

\date{Received: date / Revised version: date}

\abstract{Inspired by the bankruptcy of Lehman Brothers and its consequences
on the global financial system, we develop a simple model 
in which the Lehman default event is quantified as having an almost immediate
effect in worsening the credit worthiness of all financial institutions in the economic network.
In our stylized description, all properties of a given firm are
captured by its effective credit rating, which follows a simple dynamics
of co-evolution with the credit ratings of the other firms in our economic network.  The 
dynamics resembles the evolution of Potts spin-glass with external global field corresponding 
to a panic effect in the economy. 
The existence of a global phase transition, between paramagnetic and ferromagnetic phases, 
 explains the large susceptibility
of the system to negative shocks. We show that bailing out
the first few defaulting firms does not solve the problem, but does have the effect of alleviating
considerably the global shock, as measured by the fraction of firms that
are not defaulting as a consequence. This beneficial effect is the counterpart
of the large vulnerability of the system of coupled firms, which are both
the direct consequences of the collective self-organized
endogenous behaviors of the credit ratings
of the firms in our economic network.
\PACS{
{89.65.Gh}{Economics; econophysics, financial markets, business and management}
}
}

 \maketitle




\section{Introduction}

The largest financial crisis since the great depression started in 2007 with an initially well-defined epicenter focused on mortgage backed securities (MBS). It has since been cascading into a global economic recession, whose increasing severity and uncertain duration has led and is continuing to lead to massive losses and damage for billions of people. 
This crisis has brought to the attention of everyone the concept that risk can be endogenous and can cascade 
as a growing avalanche through the financial and economic system. This is the opposite of the assumption
held previously by many regulators, credit agencies and bankers that risks can be managed by using models
essentially focusing on the risk of each single institution, and by combining them using 
assumptions of inter-dependencies calibrated in good times. The unveiled systemic nature of financial risks
makes now clear the need for global approaches including economic and financial networks and modeling
the collective behaviors that results from the coupling between institutions, firms, financial products and so on.

During the development of the crisis in 2008, a succession of problems unfolded, the most prominent 
ones being associated with names such as Bear Stearns, Fannie, Freddie, AIG, 
Washington Mutual, and Wachovia. Except for the famous bankruptcy of Lehman Brothers Holding Inc., 
all the other ``too-big-to-fail'' financial institutions and insurance companies were bailed out,
while thousands of smaller banks have been left alone to undergo bankruptcy. 

We propose here a minimalist framework to account for the observed cascade
of defaults, which incorporates the impact of changes of beliefs in credit worthiness
of a given firm, resulting from the change of credit worthiness of other institutions
in the same economic network. In this way, we are able to account for the 
cascade phenomenon, and explain how small changes can lead to dramatic
consequences, all the more so, the more coupled is the economic network
and the larger is the number of firms. Our framework allows us to also analyze the 
so-called ``Lehman Brothers'' effect, i.e., the crash on the stock market
and the subsequent panic among financial institutions, following its 
filling for bankruptcy on 15 September 2008. We account for this aftermath
by introducing a global field influencing every firm by enlarging
the probability of rating downgrade, which embodies the psychological impact of
the destruction of trust among institutions, which suddenly realized
that, after all, the US Treasury and the Federal Reserve might not bail them
out. The abrupt drop in confidence led to a drop in credit supply
and to the major recession.
Finally, we investigate, within our set-up, the effect of a simple bail-out policy
consisting of accepting to rescue a certain number of defaulting firms.
We find that this rescue policy is the most efficient when the economy
is functioning at its most vulnerable state of critical coupling between its
constituting firms. 

Many models have been proposed to investigate the effects
of credit contagion on the default probability of individual firms.
Motivated by the accounting scandals at Enron, Worldcom and Tyco, Giesecke \cite{Giesecke04}
developed a structural model of correlated multi-firm default, in which investors
update their belief on the liabilities of remaining firms after each firm default, which leads
to contagious jumps in credit spreads of business partners. 
Eisenberg and Noe created a model of contagion in a network of liabilities \cite{EisenbergNoe}. When 
an agent cannot fulfill its obligation, it defaults and the loss propagates along the network. 
Propagation of this distress can trigger off another defaults by contagion. 
Battiston et al. \cite{Battistonetal07} studied a simple model of a production network in which firms are linked
by supply-customer relationships involving extension of trade-credit. 
They recover some stylized facts of industrial demography and the correlation, over time
and across firms, of output, growth and bankruptcies.
Delli Gatti et al. \cite{Delligattietal} studied the properties of a credit-network economy characterized by credit
relationship connecting downstream and upstream firm through trade credit
and firms and banks through bank credit. They included a change in 
the network topology over time due to an endogenous process
of partner selection in an imperfect information decisional context, which 
leads to an interplay between network evolution and business fluctuations
and bankruptcy propagation. 
The bankruptcy of one entity can bring about the bankruptcy of one
or more other agents possibly leading to avalanches of bankruptcies. 
Azizpour et al. \cite{Azizpour} developed a self-excited model of correlated event 
timing to estimate the price of correlated corporate default risk.
Sakata et al. \cite{Sakata06} introduced an infectious default and recovery model of
a large set of firms coupled through credit-debit contracts and determined
the default probability of $k$ defaults. 
Ikeda et al. \cite{Ikeda07} have developed 
an agent-based simulation of chain bankruptcy, in which a decrease of revenue by the
loss of accounts payable is modeled by an interaction term, and bankruptcy is
defined as a capital deficit. 
Lorentz et al. \cite{Lorenzetal09} have
introduced a general framework for models of cascade and contagion processes on
networks using the concept of the fragility of a firm, including the bundle model, the voter model, and models of epidemic spreading as special cases. 
Ormerod and Colbaugh \cite{Ormerod-Colbaugh} developed a model of heterogeneous agents interacting in an
network evolving according to the formation of alliances based on the decisions of 
self-interested fitness optimizers. In the presence of external negative shocks, 
they find that increasing the number of connections causes an increase in the average 
fitness of agents, and at the same time makes the system as whole more vulnerable to 
catastrophic failure/extinction events on an near-global scale.
Schafer et al. \cite{Schaferretal07} set up a structural model of credit risk for correlated portfolios containing many
credit contracts exposed to risk factors which undergo jump-diffusion processes and
derive the full loss distribution of the credit portfolios.
 Neu and K\"{u}hn \cite{Neu-Kuhn2004} generalized existing structural models for credit risk in terms
of credit contagion with feedbacks,
using the analogy to a lattice gas model from physics. 
Already for moderate micro-economic dependencies, they find that, for stronger mutually supportive
relationship between the firms, collective phenomena such as bursts and avalanches of
defaults can be observed in their model.  Hatchett and K\"{u}hn \cite{Hatchett-Kuehn} 
studied the model of \cite{Neu-Kuhn2004} and found it to be solvable 
for the loss distributions of large loan portfolios with fat tails.
 Anand and K\"{u}hn  \cite{AnandKuhn07} extend these analyzes by studying 
the functional correlation approach to operational risk and discover
the coexistence of operational and nonoperational phases, such that 
their credit risk systems are susceptible to discontinuous phase transitions from the
operational to nonoperational phase via catastrophic breakdown. 
Their model is also relevant to understand the cascades of unprecedented losses suffered
in August 6, 2007,  by a number of
high-profile and highly successful quantitative long/short equity hedge funds
\cite{Khandani-Lo}.

The organization of the paper is as follows. Section 2 describes the model,
with the definition of the key variable, the ``effective credit rating grade'' (ECRG),
its dynamics, and how we take into account the interdependencies between firms
through the changes of their ECRGs. Section 3 presents the main results
of how a global crisis occurs in our model, and the relevance of a phase transition
in the presence of the ``panic'' effect. Section 4 examines the sensitivity of the paths of 
defaulting economies as a function of the average coupling strength between
firms, the panic field and the initial probabilities of ECRG changes. Section 5
presents some results on the impact of a simple rescue policy consisting
in preventing the first few defaults to occur. Section 6 concludes.

\section{Description of the model}

Our model is adapted from  Sieczka and Hołyst \cite{Sieczka}, who introduced
a stylized model of financial contagion among interacting companies.

\subsection{Definition of the key variable: ``effective credit rating grade'' (ECRG) $R$}

Since we wish to focus on the possible contagion and cascade of
defaults among the firms in a network, we abstract from the 
multidimensional complexity of all the variables impacting a firm's financial health.
We propose to capture the solvability of a given firm $i$
by a single variable $R_i$ that we refer to as the ``effective 
credit rating grade'' (ECRG). In our simulations, we
will consider a rather small number of effective credit rating grades, 
that is, we assume that the $R_i$'s take
discrete values: $0, 1, ..., R_{\rm max}$, with $R_{\rm max}=7$.
When $R_i(t) = 7$, investors (who are international to the networked economy) consider agent 
$i$ to be a safe investment. As the grade decreases, the risk perception of the 
investors on agent $i$'s future profitability (and viability) increases. In other words, 
investors are more skeptical of agent $i$ and hence less likely to lend to agent $i$ in 
the future. For $R_i(t) = 0$, agent $i$ has defaulted on at 
least one of its obligations to the lenders. This defaulted state is 
assumed to be an absorbing state 
in that, once an agent has defaulted in the economy, it remains 
so for the remainder of  time. The effective credit rating grade $R_i$
associated with each firm provide
an integrated, low dimensional  and effective indicator of 
the investors' subjective perception of the creditworthiness of  firm $i$.
We conjecture that 
the effective credit rating grade $R_i$ can be obtained as the end
result of a Mori-Zwanzig projection technique applied to 
a full behavioral model, after integrating over all degrees of freedom
except these effective ratings degrees of freedom, similarly to 
the approach of Neu and Kuhn on market risks  \cite{KuhnNeu08}.


The term ``effective credit rating grade'' derives obviously from 
the standard credit rating which is provided by 
credit rating agencies such as Fitch Ratings, Moody's Investors Service or
Standard \& Poor's in the U.S.  Indeed, it is well-known that  
there is a strong correlation between credit quality and
default remoteness: the higher the rating,
the lower the probability of default, and
the lower ratings always correspond
to higher default ratios \cite{BrandBahar2001}. 
With default as our target, our model can also be thought of
as a description of the world of financial instruments and their
interactions. Thus, our model can also be taken as a description of
a network of various financial instruments, including collateral debt
obligations and structured asset-backed security, which formed
the substrate on which the 2007 crisis developed.
Note that the choice of the
discrete values, $0, 1, ..., R_{\rm max}$, with $R_{\rm max}=7$, for $R_i$
is motivated by the principal grades (AAA, AA, A, BBB, BB, B, CCC)
used by rating agencies. 
We associate deterministically the eighth lowest level $R=0$ to the 
state of default. This corresponds to a stylized
representation of the well-known abrupt increase of default rates as credit rating
deteriorates. For instance, for the period from 1981 to 1999,
Standard \& Poor's reported a $0.00\%$ probability
of default  per year for AAA-rated firms, compared with a value of $5.3\%$ per year
for B-rated firms and a value of $22\%$ for CCC-rated firms \cite{BrandBahar2001}. In the same vein,
the probability of a default over a 15 year period is $0.5\%$ for AAA-rated firms,
compared with $30\%$ for B-rated firms and of $48.3\%$ for CCC-rated firms.
 
Our aim is to account for the many complexities associated with the default hazard of
each firm by focusing on the dynamics of a single variable for each firm.
Then, the question arises as whether the dynamics of the standard credit ratings are 
faithful indicators of the evolution of firm default hazards. The answer
is probably negative, in view of the wave of defaults that started in 2007
and accelerated in 2008, in which many financial instruments, special
vehicles as well as major insurance companies and investment banks that were rated 
AAA turned out to default. 

This is the principal reason for our proposal to think of each $R_i$, as not being
the official published credit rating grade but, as an
``effective credit rating grade'', which can be interpreted in several ways.
It could be the real internal credit worthiness, 
that credit agencies should strive to uncover. 
Another way to think about the effective credit rating grade is that it
is the rating that the rating agency has produced internally but not yet published
due to their well documented conflict of interest leading to 
patently misleading published ratings.
Indeed, it is now well-recognized
that the rating grade provided by rating agencies are imperfect, as 
shown by the development of the financial crisis since 2007:
(i) credit rating agencies do not downgrade companies promptly enough;
(ii) credit rating agencies have made significant errors of judgment in rating structured products, 
particularly in assigning AAA ratings to structured debt, which in a large number of cases 
has subsequently been downgraded or defaulted. 
The ``effective credit rating grade'' of a given firm
can be also interpreted as quantifying the risk perception of the market
concerning that firm.

\subsection{Dynamical evolution of the $R_i$'s, for $i=1,...,N$ in a network of $N$ firms}

We consider a network of $N$ firms operating in a common economic environment. 
We assume that the dynamics of the financial 
health of firms can be captured by that of their effective credit rating grades (ECRG) $R_i(t)$'s.
We assume the simple discrete correlated random walk with constraints, 
\begin{equation}
 R_i(t)=R_i(t-1)+s_i(t),
\label{eq:R}
\end{equation}
where $s_i(t)$ is a stochastic variable taking only 
three possible values $s_i(t)=-1, 0, 1$.

This means that the ECRG of a given firm can change 
by no more than one level over one time step, i.e., $|R_i(t)-R_i(t-1)|\leq 1$. 

Since a value of the ECRG of a firm equal to $7$ corresponds to the maximum possible level,
if $R_i(t-1)=7$, then $R_i(t)$ can only either remain at the same level or decrease by $1$.
This means that the one--time-step ECRG change $s_i(t)$ of a firm can only take values $0$ or $-1$
if $R_i(t-1)=7$. This corresponds to a ``reflecting'' condition of the discrete correlated random
walk (\ref{eq:R}). 

We assume that, when a firm or a financial product defaults, it does not recover and remains
at the default state $R_i=0$. This corresponds to $s_i$ remaining equal to $0$ for all
subsequent time steps and this is an ``absorbing'' condition for the discrete correlated random
walk (\ref{eq:R}).  Because we are describing a short time span of just a few years in our
simulations, we do not include a firm entry process. Neglecting the creation of new 
banks seems to be a reasonable assumption when thinking of the
economy of banks during the financial crisis.

\subsection{Interdependencies between firms in the changes of their effective credit rating grades}

There is rich literature on networks of firms, including credit and corporate ownership networks,
as well as production, trade, supply chain and innovation networks (see e.g. \cite{SchweitzerScience09,SchweitzerACS09} 
and references therein). These links between firms are concrete and quantifiable through cash flows
or control structures (such as voting rights). But, many of these links, forming 
complex networks of creditor/obligor relationships, revolving
credit agreements and so on, are largely unmapped \cite{Lo08}. This supports 
considering another type of networks, 
a network of more intangible but nonetheless essential relationships 
of how the credit rating grades of different firms interact.
This idea is also suggested by the cases of the Asian crises of 1997 
and the more recent subprime crisis of 2007-2008.

Consider first the case of the Asian crises in 1997 and the remarkable
fact emphasized by Krugman \cite{Krugman_depressioneconomics08}
 that the traditional measures of vulnerability did not forecast the crisis.
One possible explanation is that the problem was off the
government's balance sheet, and not part of the governments'
visible liabilities until after the fact. The crisis developed 
as a self-fulfilling generated downward spiral of asset
deflation and disintermediation. Post-mortem 
analyses have generally concluded that 
the key ingredient was that many of the Asian countries at that time had 
either pegged their currency to the dollar or operated in a narrow exchange rate band. 
In addition, asset price bubbles were growing in these countries,
catalyzed by an influx of foreign funds. With deteriorating economic conditions, 
these economies became ripe for speculative attacks and bank runs. 
The sequence of events was one of 
speculative attacks on the currency pegs in Thailand and Malaysia followed by flight of 
capital away from these and surrounding economies,  a process 
sometimes referred to as twin-crisis \cite{Goldsteintwincrises05}.

This all looks clear in hindsight, but the notion of ripeness to speculative
attacks and bank runs is a matter of debate. For instance, no bank, however sound its
capital and liability structure, can survive a bank run, by construction.
Krugman \cite{Krugman_depressioneconomics08} emphasizes
another important effect, what we could call a ``virtual'' network effect:
while the real economic and financial links between the strongest
Asian tigers (South Korea and Hong-Kong) and the others 
were weak on a relative GDP measure, foreign investors lumped them 
in their mental framework as part of the same ``Asian'' portfolio.
Starting in Malaysia in July 1997,
the contagion of the crisis to the other Asian countries was actually strongly amplified
by the misperception by foreign investors that these different
economies (Thailand, Malaysia, South Korea, Hong Kong...)
were strongly linked. According to this mental framing \cite{Ariely09}, the problem of Thailand
was not just a one-country happenance due to a localized over-indebtness,
but it was thought that this was the problem of the whole economic Asian zone.
While the vulnerability of South Korea and Hong Kong has been disputed,  the incorrect geographical
emphasis made the links become self-created by the western investors and banks
pulling out from all these countries simultaneously in a process
analogously to a bank run (see the argument developed 
in some details in \cite{Krugman_depressioneconomics08}).

The causes of the financial and economic crisis initiated by 
the subprime crisis starting in 2007 are multiple and intertwined. The following
elements have been documented and argued to be important
contributors: (i) Real-estate loans and MBS (morgage backed securities) as fraction of bank assets,
leading to feedback loops in leverage and ultimately in fragility; (ii) Managers' greed and poor corporate governance problems;
(iii) Deregulation and lack of oversight; (iv) Bad quantitative risk models in banks (Basel II);
(v) Lowering of lending standards; (vi) Securitization of finance; (vii) Leverage; 
(viii) Rating agency failures; (ix) Under-estimating aggregate risks; (x) Growth of over-capacity;
(xi) the agenda of several successive US administration to promote accession to house ownership
to the middle-class and poor; (xii) Government sponsored entities such as Freddy Mac and Fanny Mae
with unfair access to liquidity and mispriced returns due to the implicit Government put option.
There is an enormous contemporary literature, which is growing everyday and which analyzes
these different factors and suggests new solutions and new designs. 

However, Sornette and Woodard argued recently \cite{SorWood10} that these elements
were part of and contributed to a more global process, dubbed ``the illusion of the perpetual 
money machine,'' that developed over the last twenty years. 
During its development, the shadow banking
(with a capital value peak of \$2.2 trillions in early 2007) constituted
a large network of inter-dependencies between financial 
instruments and between investment houses including banks, 
that became apparent only after the crisis started to unfold \cite{Krugman_depressioneconomics08}.
This shadow banking system included such
financial instruments as the 
auction rate preferred securities, asset-backed commercial paper,
structured investment vehicles, tender option bonds and
variable rate demand notes. These different instruments are indeed
interacting directly through their overlapping collaterals and underlying assets, 
and indirectly through their impacts onto investment portfolios,
through the joint impact of economic and/or financial shocks and, 
very importantly through the psychological contagion of investors.

These elements motivate us in modeling the 
direct as well as indirect couplings between firms through
the effective credit rating grade changes $s_i$'s. This provides us with a coarse-grained
description that has the advantage of lumping together many 
coupled mechanisms.  Following the theory of multinomial choice \cite{McFadden74,McFadden78}, we formulate
the conditional probability for the ECRG change $s_i(t)$ of a given firm $i$ under the form
\begin{equation}
\label{eq:Ps}
\begin{split}
 P&(s_i \vert s_1, ..., s_{i-1}, s_{i+1}, ..., s_N, R_1,..., R_N)\\
&=\frac{1}{Z}\exp\left(\sum_{j\neq i}J_{ij}\delta_{s_i s_j}+ H(R_1,..., R_N)\delta_{s_i -1}\right)~,
\end{split}
\end{equation}
where $\delta_{s_i s_j}$ is the Kronecker delta and $Z$ ensures a proper normalization. 
Here, the $s_i$ on the left side of $\vert$ is realized at time $t$, while the 
conditioning variables on the right side of $\vert$ are realized at time $t-1$. 
Expression (\ref{eq:Ps}) gives the dependence of the probability that the ECRG change $s_i(t)$ of a firm $i$ 
takes a value $+1$, $0$ or $-1$, given the previous ECRG changes $s_j(t-1), j \neq i$ at $t-1$,
and given the ECRG level $R_1(t-1), ..., R_N(t-1)$ at time $t-1$. The probability of a rating change does not 
depend on a rating history. It is an assumption that agrees with a behavior of the Merton model in which a value
 of a firm's asset evolves as a geometric Brownian motion and rating classes can be translated into thresholds of
the firm's asset value.

This Ising-like expression (\ref{eq:Ps}) is also motivated by the demonstration that 
finite-size long-range Ising model turns out to be
an adequate model for the description of homogeneous credit portfolios and
the computation of credit risk when default correlations between the borrowers are included
\cite{MolinsVives}.
This probability (\ref{eq:Ps}) is controlled by
two contributions. 
\begin{enumerate}
\item The term $\sum_{j\neq i}J_{ij}\delta_{s_i s_j}$ describes a mimetic or contagion effect, 
as well as a persistence mechanism: the larger the number of firms which have seen their grading updated upward
(respectively unchanged or downward), the more new firms will have their grading updated
upward (respectively unchanged or downward) in the next time step.

The strength of the influence of firm $i$ on firm $j$ and vice-versa (as we assume for simplicity
symmetric couplings) is quantified by the $N \times N$ symmetric interaction matrix  $\{J_{ij}\}$. 
The elements $J_{ij}$ can take zero, negative, or positive values. For $J_{ij}=0$, 
the two firms do not influence each other through the contagion effect. 
A positive $J_{ij}$ corresponds to two dependent firms in the same 
industry branch, such that an improving (respectively worsening) condition
in one of them tends to improve (respectively worsen) the situation of the other firm.
In contrast, a negative $J_{ij}$ may describe anti-correlations, such that 
one firm might profit at the expanse of the other one. As there is clear evidence
of a strong bias toward positive correlations among firms, we assume that the $J_{ij}$'s are normally distributed 
with a positive mean $J_0$ and standard deviation $\sigma_J$:
\begin{equation}
\label{ghy3bg}
 P(J_{ij})=\frac{1}{\sqrt{2\pi \sigma_J^2}}\exp \left(-\frac{(J_{ij}-J_0)^2}{2\sigma_J^2}\right)~.
\end{equation}

\item The term $H(R_1,..., R_N)\delta_{s_i,-1}$
embodies the sentiment of panic, which was triggered when Lehman Brothers Holding Inc., 
a global financial-services firm, declared bankruptcy on 15 September 2008. The field 
$H(R_1,..., R_N)\delta_{s_i,-1}$ is equal to zero 
until the first ban\-krupt\-cy occurs in the network of firms, after which it remains fixed to 
$H\delta_{s_i, -1}$: after 
the first bankruptcy occurs in our network, all credit rating grades are biased towards
deteriorating, where $H$ is the strength of this bias. Formally,
\begin{equation}
\label{eq:H}
 H(R_1,..., R_N)=H\cdot (1-\prod_j (1-\delta_{R_j, 0})),
\end{equation}
which is equal to $H$ when at least one firm defaulted ($R_i=0$), otherwise the field is equal to $0$.  
The field $H\delta_{s_i, -1}$ models a global shock which appears after the default of a first company.  
 The ``panic'' field $H$ captures the psychological impact of the destruction of trust among
institutions. Its structural impact is reasonable in the interpretation of the 
``effective credit rating grade'' of a given firm
as quantifying the risk perception of the market concerning that firm. The activation of $H$
describes the contagion effect of the increased risk perception. 
The activation of $H$ can also be justified as describing a genuine loss of 
real internal credit worthiness following a shock on one of the firms, due to the
inter-connected liabilities in the bank balance sheets. 

Finally, the
panic term in equation (\ref{eq:Ps}) is either $H$ (if there is at 
least one bankrupted firm) or $0$ (if there is none). It could be argued 
to be more realistic to make the panic term depending on the number of 
bankrupted firms. Our simplifying framework consists in considering only 
the systemically important institutions such as 
Bear Stearns, Fannie, Freddie, AIG, 
Washington Mutual, Wachovia and so on, which
are considered consensually to be ``too-big-to-fail'' financial institutions and insurance companies.
Just one of them going to bankruptcy is arguably sufficient to trigger
the Lehman Brothers type of panics that we model here.
\end{enumerate}

Our goal is to study what is the effect on the global firm network of the 
change in psychological attitude of investors following the first bankruptcy.

\section{Global crisis as a phase transition in the presence of the ``panic effect''}

\subsection{Description of the numerical simulation procedure \label{th3bw}}

We simulated a system of $N=1000$ firms. The standard deviation of the coupling coefficients $J_{ij}$
defined by (\ref{ghy3bg}) is set to $\sigma_J=0.001$. We fix the average value $J_0$ of the coupling
coefficients and the value $H$ of the ``panic field.''
\begin{enumerate}
\item We generate a given realization of a randomly initialized matrix $J_{ij}$;
 \item The initial sets of $\{s_i\}$ and $\{R_i\}$ are generated from uniform distributions (with exclusion of $R=0$);
 \item A randomly selected spin $s_i$ is updated according to (\ref{eq:Ps}), followed by $R_i$ which is updated according to (\ref{eq:R});
\item We repeat the previous step $N$ times in total, so as to update the rating of $N$ firms, which defines one complete time step;
\item We then go back to steps 3 and 4 and iterate them $8$ times, which corresponds
to a total lifetime of $8$ time steps;
\item We count the number of defaults ($ND$) that have occurred over these $8$ time steps;
\item The whole procedure starting from step 1 to step 6 is repeated 1000 times to generate 1000 different realizations over which we can average over the random realizations of the matrix $\{J_{ij}\}$ and 
over the initial values of $\{s_i\}$ and $\{R_i\}$.
\end{enumerate}
In the following, we analyze different properties of the systems of $N$ firms as a function of
$J_0$ and $H$.

\subsection{Critical point and susceptibility}


\subsubsection{Absence of the global ``panic'' field ($H=0$) \label{thnbw}}

In the absence of the panic field ($H=0$), the dynamics of the $s_i$'s
defined by model (\ref{eq:Ps}) with the updating rules described in subsection \ref{th3bw}
is nothing but the implementation of the Glauber algorithm \cite{Glauber63} applied to the Potts model 
for a fully connected Potts model with random couplings $\{J_{ij}\}$ with strictly positive average 
${\rm E}[J_{ij}] := J_0 >0$.
For such a system, it is known \cite{SherringtonKirkpatrick,GabayToulouse,Nishimori} 
that, in the limit $N \to +\infty$, a phase transition
separates a (so-called paramagnetic) phase at $J_0 < J_{0c}$ where the $s_i$'s on the average occupy all three states in equal numbers,  from a (so-called ferromagnetic) phase at $J_0 > J_{0c}$ 
where one state of the $s_i$'s  dominates over the remaining two.
If we define $p$ and $q$ as the relative numbers of $s_i$'s in states $-1$ and $1$ respectively, we can describe the system using the pair $(p,q)$. The system in the paramagnetic phase is characterized by $(1/3, 1/3)$. In case of a ferromagnetic phase, three equi-probable pairs $(p_i(J_0, \sigma_J), q_i(J_0, \sigma_J))$ with $p_i\neq 1/3$, $q_i\neq1/3$ are possible and the system selects one of them randomly according to the influence of initial conditions and its noisy dynamics. This is the phenomenon of ``spontaneous symmetry breaking'', 
defined as the situation in which the solution selected by the dynamics has a lower symmetry
than its equation. The concept of spontaneous symmetry breaking has many important applications
in many fields of science \cite{Aravind,Consoli,Drugowich,Goldenfeld92,Sivardiere1,Sivardiere2,Weinberg},
with deep implications that span the creation of the universe,
the fundamental interactions in physics, the emergence
of particle masses to the emergence of stock market speculation \cite{Sornettespec2000}.
When the average coupling strength $J_0$ is equal to the critical value $J_{0c}$,
the system is at its critical point at which its susceptibility diverges for $N \to \infty$
(and grows as a positive power of $N$ for finite $N$).  The numerical value of $J_{0c}$
in the limit $N \to +\infty$ is a well-defined function of $\sigma_J$  \cite{Nishimori}.

\begin{figure}
 \centering
 \includegraphics[scale=0.5, angle=-90]{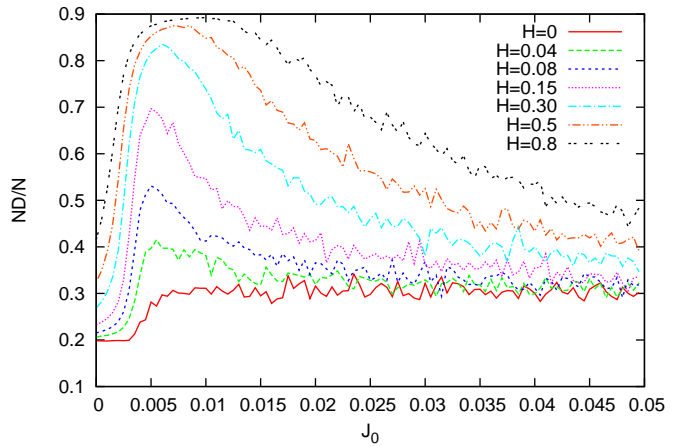}
 \caption{Relative number of defaults (ND/N) as a function of $J_0$ for different values of the ``panic'' field amplitude $H$.}
 \label{fig:1}
\end{figure}

For $H=0$, we thus expect each ECRG $R_i$ to follow a kind of correlated
{\it unbiased} random walk for $J_0 < J_{0c}$. In contrast, for $J_0 > J_{0c}$, 
the spontaneous symmetry breaking of the average of the $s_i$'s translates into 
a non-zero bias. With probability $1/3$, this bias is positive, tending to push
the credit ratings upwards. With probability $1/3$, the bias is negative, pushing
the firms inexorably towards the default state $R=0$. And with probability $1/3$, the bias is neutral, keeping the credit rating in its position.  
Figure \ref{fig:1} presents the average cumulative number of defaults observed in our 
simulations over $8$ time steps, averaged over 1000 realizations, as a function of $J_0$ for different field amplitudes $H$. 
For $H=0$, we observe that the fraction of firms that have defaulted in the time span
of $8$ time steps jumps from $\simeq 0.2$ to $\simeq 0.3$ as $J_0$
passes through a critical value $J_{0c} \approx 0.0045$. This is the signature of the
bias just mentioned above associated with the spontaneous symmetry breaking
occurring above the critical point at $J_{0c}$. 


\begin{figure}
 \centering
 \includegraphics[angle=-90, scale=0.7]{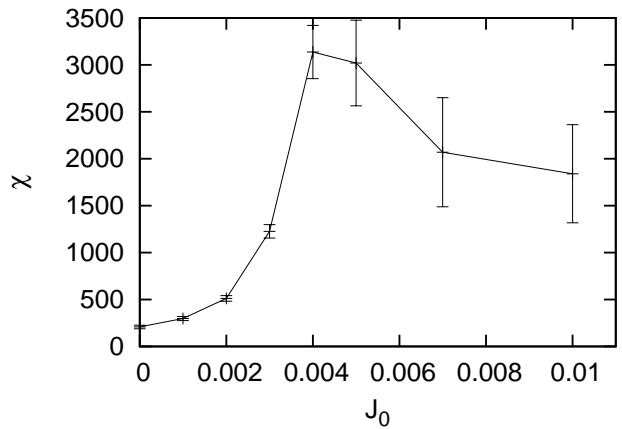}
 \caption{Susceptibility $\chi(J_0)$ defined by (\protect\ref{yhy3ybr}) for $N=1000$ with errorbars corresponding
to a standard deviation.}
 \label{fig:2}
\end{figure}

\subsubsection{Effect of the global ``panic'' field ($H>0$)}

For $H>0$, the number of defaults $ND$ develops a maximum for $J_0$ in the vicinity of the critical point $J_{0c}$
and the value of the maximum is higher for higher values of the panic field $H$. This peak
is due to the phenomenon mentioned above that the neighborhood of the critical point
is associated with an increased susceptibility of the system to external perturbations.
In the present context, the susceptibility (or we should rather say the vulnerability to panic)
can be defined as
\begin{equation}
 \chi=\frac{\partial ND}{\partial H}\Big | _{H=0}~.
 \label{yhy3ybr}
\end{equation}
A numerical estimation of the susceptibility for the system of $N=1000$ nodes and different $J_0$ is presented in Fig. \ref{fig:2}. A maximum of $\chi$ near the critical point is clearly visible. The peak is expected to diverge
as the system size $N$ goes to infinity.

\begin{figure}
 \centering
 \includegraphics[scale=0.45, angle=-90]{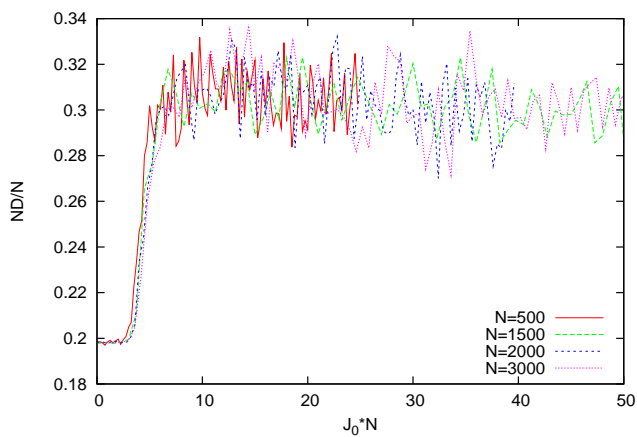}
 \caption{Dependence of $ND$ scaled by the system size $N$ versus $J_0$ in the absence of the field ($H=0$) for different system sizes $N$. }
 \label{fig:3}
\end{figure}

\begin{figure}
 \centering
 \includegraphics[scale=0.45, angle=-90]{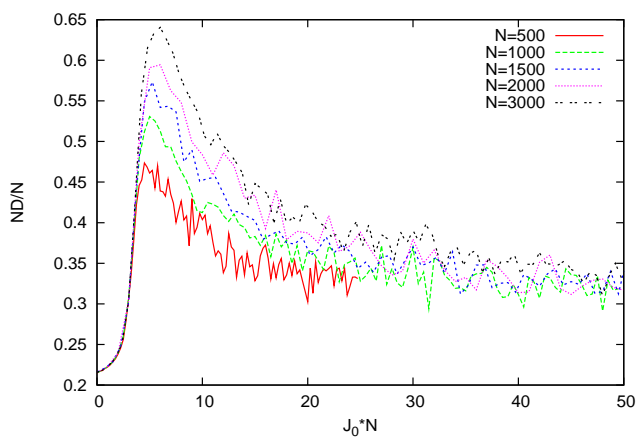}
 \caption{Same as Figure \protect\ref{fig:3} for $H=0.08$.}
 \label{fig:4}
\end{figure}

\subsubsection{Critical susceptibility and non-monotonous nonlinear amplification of negative shocks}

The interplay between the panic field $H$ and the collective phenomenon of defaults
is well illustrated by comparing Figure \ref{fig:3} and Figure \ref{fig:4}.
Figure \ref{fig:3} for $H=0$ shows that the number of defaults scaled by the size
$N$ of the economy is independent of $N$. In other words, for all values
of the average coupling strength $J_0$, the number of defaults is simply
proportional to the size of the economy all along the nonlinear path as a function
of $J_0$. This result illustrates that the severity of the crisis which,
in an economy where firms are coupled to each other, does not remain localized
to a few firms but scales with the size of the economy.

Figure \ref{fig:4} shows that, in the presence of the panic field $H$, the simple
scaling of Figure \ref{fig:3} does not hold anymore. It is replaced by a strong nonlinear 
amplification of the number of defaults with the size of the economy: the scaled
variable $ND/N$ grows with growing $N$ at the peak
occurring close at the critical point $J_{0c}$. Far away from the critical point,
one can observe again a good collapse of the curves for different economy sizes.
Thus,  the nonlinear amplification as a function of system size $N$ 
is specifically associated with the behavior of the susceptibility
which tends to diverge at $J_{0c}$ as the system size $N$ increases. 

Figures  \ref{fig:1}-\ref{fig:4} carry two vivid messages.
\begin{itemize}
\item Even in the absence of panic ($H=0$), as the strength of the interactions between firms increases, 
represented by a rising $J_0$, the severity of the crisis quantified by the fraction of 
firms defaulting on their obligations exhibits an abrupt nonlinear jump around a critical value
$J_{0c}$. 

\item In the presence of the panic field ($H>0$), the situation worsens considerably,
with nonlinear and non-monotonous responses as a function of the average coupling 
strength $J_0$. The observed maximum susceptibility to a panic effect 
is associated with a very strong dependence on the size of the system:
as shown in Figure \ref{fig:4}, the larger the number of interacting firms and of coupled financial instruments in 
the economy, the much larger is the amplitude of the crisis as measured 
by the fraction of default.
\end{itemize}

Students of collective phenomena will
not be surprised by these strong nonlinearities that emerge
endogenously from the interactions between the $N$ firms
and might even take advantage of the existence of precursors
associated with the approach to the critical point (see chapter 10 on 
``Transitions, bifurcations and precursors'' of \cite{SornetteCritbook06}).
In other words, while the severity of the crisis is many times amplified in the presence 
of a panic field in a neighborhood of the critical point, making them
akin to so-called predictable  ``dragon-king'' \cite{SornetteDragon09}, the good news is that 
such behavior can be anticipated \cite{Sornettepredic2002,Scheffer-Nature09}.
The analysis presented by Sornette and Woodard \cite{SorWood10} 
supports this claim, based on the realization that
the fundamental cause of the 
unfolding financial and economic crisis stems from the accumulation of five
bubbles and their interplay and mutual reinforcement. In this vein, one of us and his 
group at ETH Zurich has started the ``financial bubble experiment'' to test
systematically and rigorously the hypothesis that financial crises can be 
diagnosed in advance \cite{FCO-FBE}. The present model provides 
a possible supporting mechanism.

\section{Sensitivity study of the paths of defaulting economies}

The evolution of firms depends on the initial condition of the economy, represented
by the set of $\{s_i\}$ and $\{R_i\}$ values at the initial time. In the simulations
whose results have been presented in Figures  \ref{fig:1}-\ref{fig:4}, we used
initial conditions for which the $3$ states $s$ and the $7$ non-zero rating level $R$ 
are equally populated. Defining  $p(t)$ and $q(t)$ as the probabilities for $s=-1$ and $s=1$ respectively
at time $t$,  this corresponds to $p(0)=q(0)=1/3$.

We now study how the evolution of the economy depends on different initial conditions for
$p(0)$ and $q(0)$. 

\subsection{Evolution of the ECRG changes by homogenized equations}

First, we study the evolution of the ECRG change $s_i(t)$
via their probabilities. The  evolution of the probability of the different states $\{s_i\}$, in absence of default, 
can be described by the following equations \cite{Sieczka}
\begin{equation}
\begin{split}
&p(t+1)=\\
&\frac{\exp(\tilde J_0p(t)+H)}{\exp(\tilde J_0p(t)+H)+\exp(\tilde J_0q(t))+\exp(\tilde J_0(1-p(t)-q(t)))},
\end{split}
\label{eq:mf_p}
\end{equation}
\begin{equation}
\begin{split}
& q(t+1)=\\
&\frac{\exp(\tilde J_0q(t))}{\exp(\tilde J_0p(t)+H)+\exp(\tilde J_0q(t))+\exp(\tilde J_0(1-p(t)-q(t)))}~,
\end{split}
\label{eq:mf_q}
\end{equation}
where ${\tilde J_0} =J_0 N$. Of course, the probability that the ECRG change is $0$ at time $t$
is simply $1-p(t)-q(t)$. These equations (\ref{eq:mf_p}-\ref{eq:mf_q}) are obtained
by using a representative firm approach, known in mathematics as an homogenization procedure,
and in physics as a mean-field approximation.

\begin{figure*}
\begin{center}
\includegraphics[angle=-90, scale=0.4]{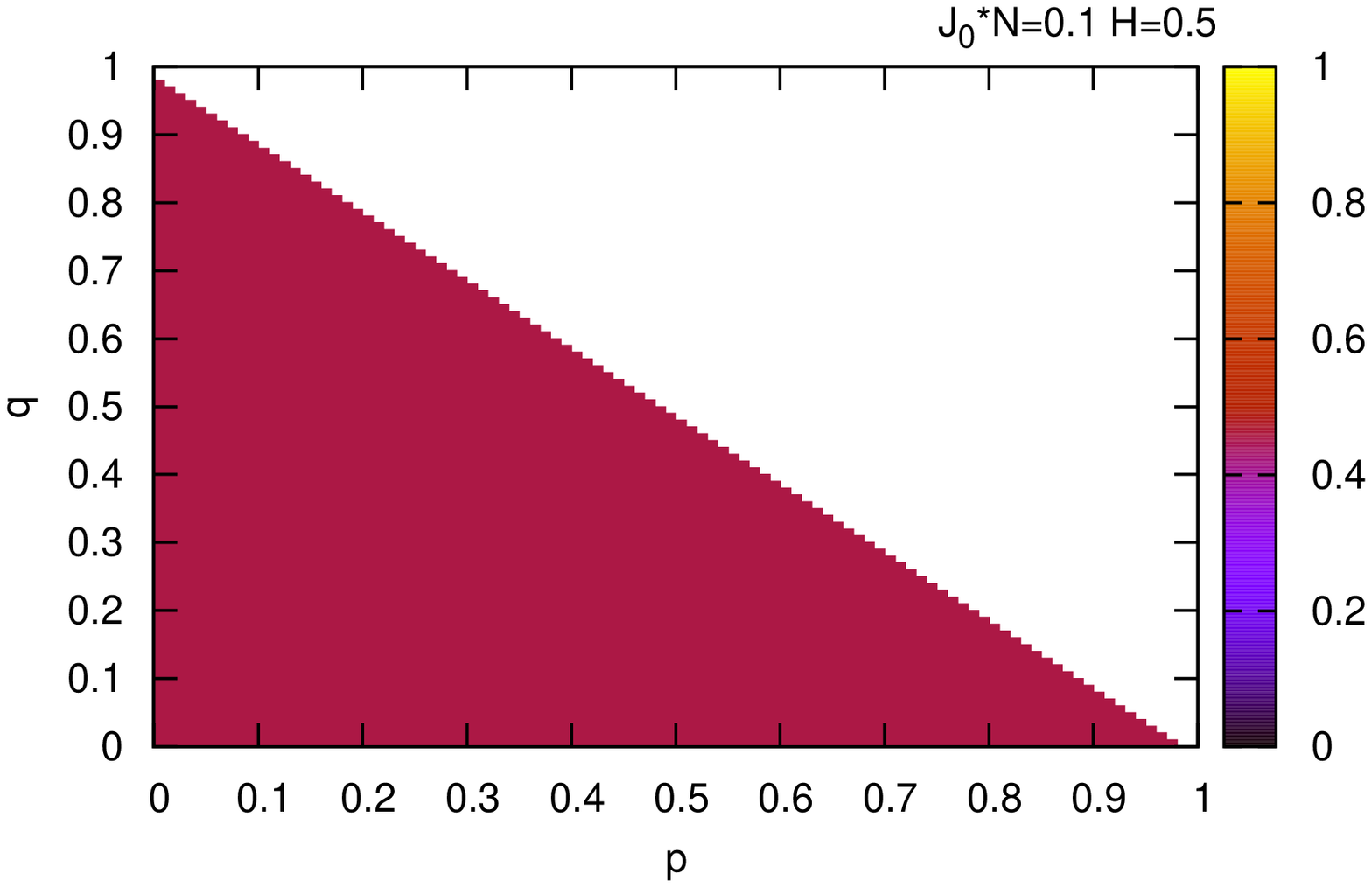} \includegraphics[angle=-90, scale=0.4]{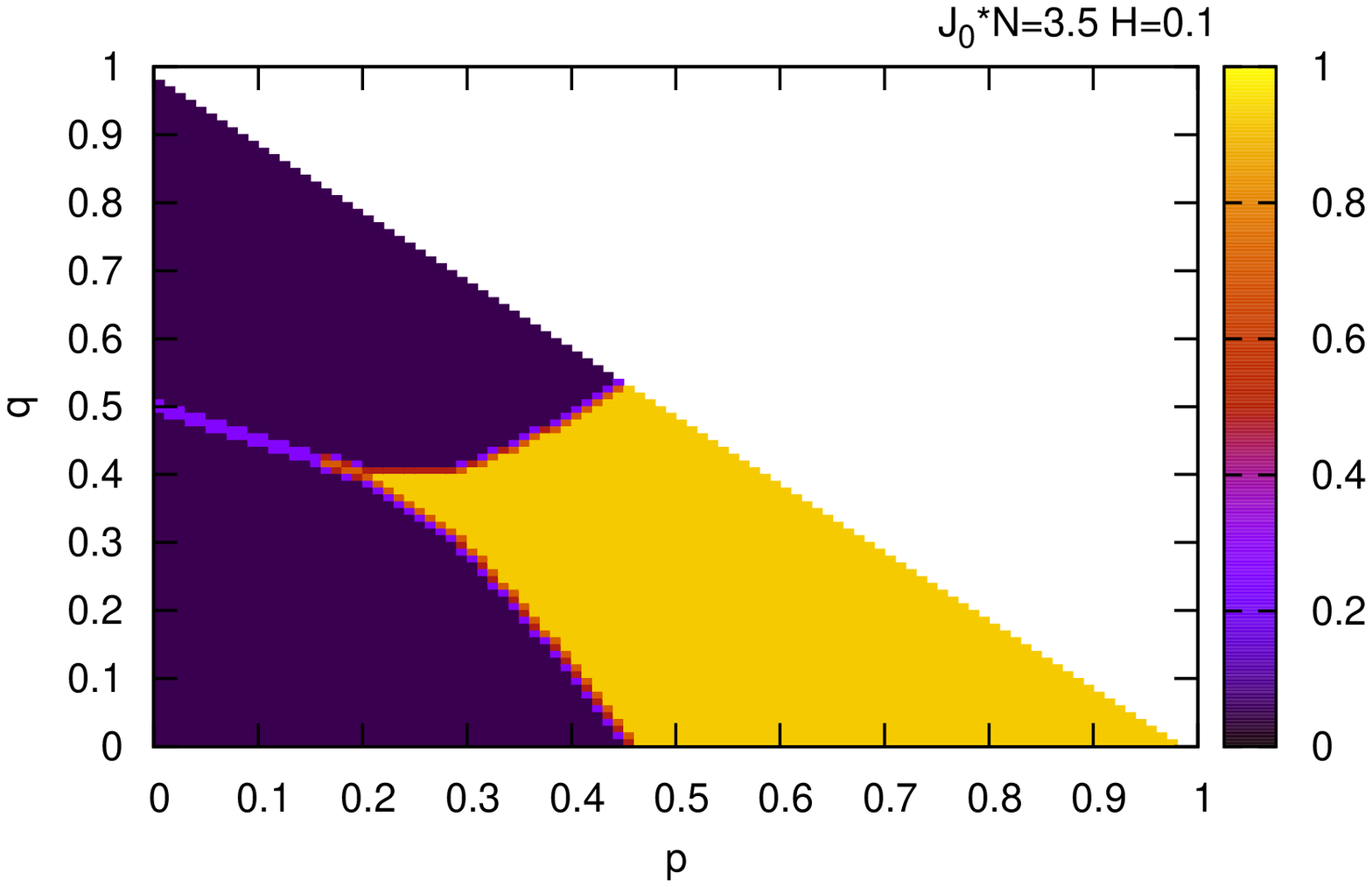}\\
\includegraphics[angle=-90, scale=0.4]{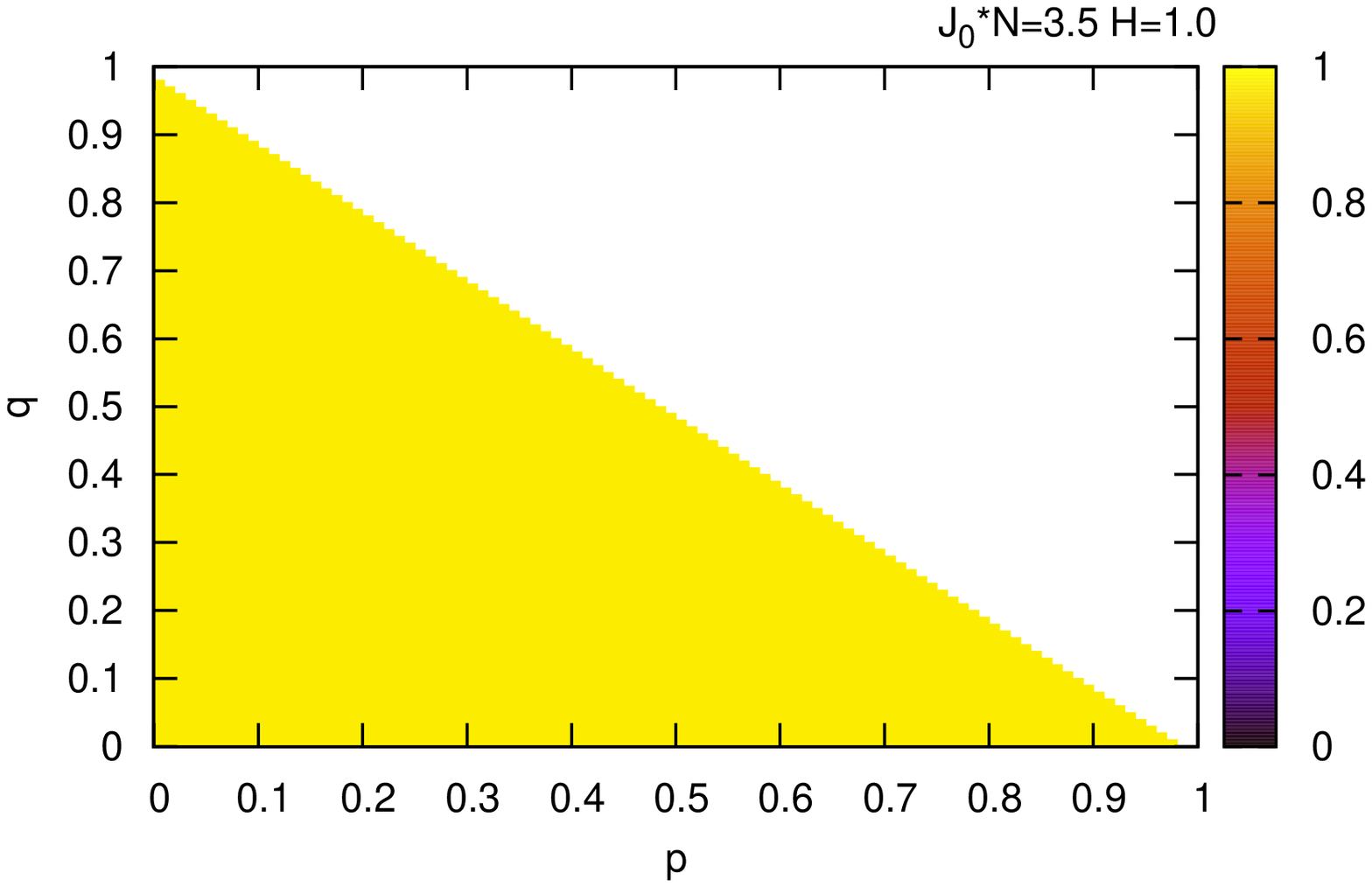} \includegraphics[angle=-90, scale=0.4]{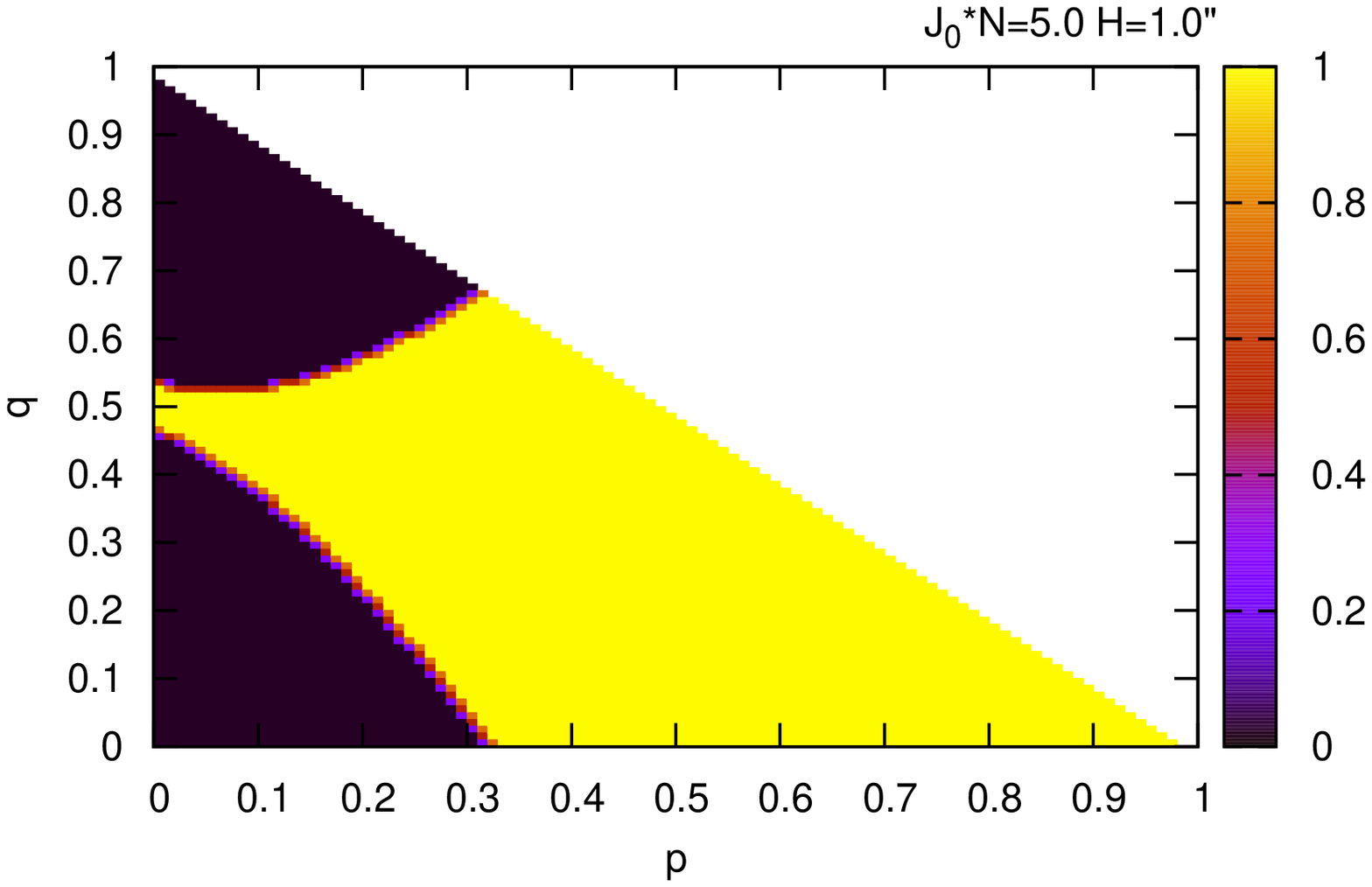}                                                                                                 
 \end{center}
 \caption{Color coding of the long-time probability $p(\infty)$ for the ECRG change of a firm in absence of default
 to take the value $s=-1$. The initial state $(p(0),q(0))$ is represented by a point on a $x, y$ plane, while the final state $p(\infty)$ is visualized by a specific color, whose value is read from the code-bar on the right of the panels. 
Upper left panel: a biased paramagnetic phase (${\tilde J_0}=0.1$, $H=0.5$); upper right panel: a ferromagnetic phase with slightly broken symmetry (${\tilde J_0}=3.5$, $H=0.1$); lower left panel: a ferromagnetic phase with completely broken symmetry (${\tilde J_0}=3.5$, $H=1.0$); lower right panel: a ferromagnetic phase with broken symmetry (${\tilde J_0}=5.0$ $H=1.0$).}
\label{fig:mf_iter}
\end{figure*}

Starting with an initial condition $(p(0), q(0))$, the equations (\ref{eq:mf_p}-\ref{eq:mf_q}) were iterated to obtain an asymptotic value $p(\infty)$ at long times.  
The results are presented in Fig. \ref{fig:mf_iter}. The initial state $(p(0),q(0))$ is represented by a point on a $x, y$ plane, while the final state $p(\infty)$ is visualized by a specific color.  It can be seen that the effect of initial initial conditions is different for different parameters $J_0$ and $H$. 

In a paramagnetic phase (${\tilde J_0} =0.1$, $H=0.5$), all initial conditions lead to the same  final state, which is biased by the field $H$. A ferromagnetic phase (${\tilde J_0} =3.5$, $H=0.1$) breaks into two final states of a complete ordering which depends on the initial condition. In the illustrated case, the symmetry is broken by the field and, as a consequence, the state $p(\infty)\simeq0.93$ is favored. 
In the ferromagnetic phase with a strong field (${\tilde J_0} =3.5$, $H=1.0$), the symmetry is completely broken and all initial conditions lead to $p(\infty)\simeq 0.98$. For higher interaction strength (${\tilde J_0} =5.0$, $H=1.0$), the system is back to a two-state competition with $p(\infty)\simeq1$ state preferred more often, due to the effect of the field. In the cases (${\tilde J_0} =3.5$, $H=0.1$) and (${\tilde J_0} =5.0$, $H=1.0$), the bottom dark regions
correspond to states ordered along the neutral axis with zero average magnetization. This rating status quo
could result from economic stagnation with no firm default but no improvement in credit ratings.

A mean field approach for an infinite range model provides the exact solution. However, applying 
a mean field approach to our model constitutes
an approximation for two reasons. First, we assume in equations (\ref{eq:mf_p}-\ref{eq:mf_q}) that the field H is nonzero from the 
beginning of the iteration which is not true in the simulations where the field appears after the first default
 happens. Second, the simulations run over a finite time, so that the asymptotic state $p(\infty)$ can only be an approximation.
 Results of the iteration presented in Fig. \ref{fig:mf_iter}
 play an illustrative role and show areas of initial conditions that lead to common asymptotic solution.

\subsection{Sensitivity of defaults on initial conditions}

Is the development of a financial crisis through a cascade of default ineluctable?
What is the role of the initial conditions and of random shocks during its development?
We can address these questions by using our model and study the impact
of different values $p(0), q(0)$ of the probabilities of the initial ECRG changes $-1$ and $+1$
for various $J_0$ and $H$.

\begin{figure}
 \centering
 \includegraphics[scale=0.4, angle=-90]{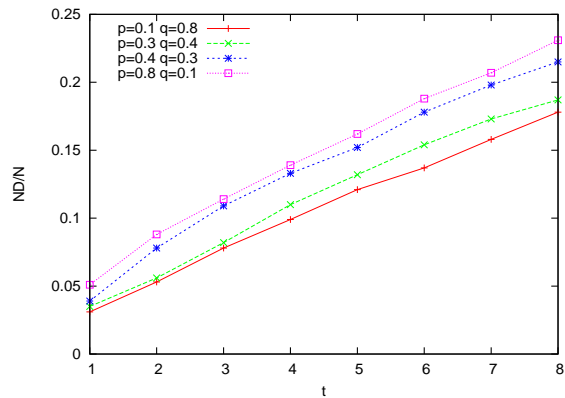}
 \caption{Evolution of the cumulative number $ND(t)$ of defaults as a function of time for 8 time steps of simulations for $J_0=0.0005$, $\sigma_J=0.001$, corresponding to the ``paramagnetic'' phase in which the average ECRG changes is zero. One can observe that different initial conditions lead to similar evolutions. The curves are obtained by averaging over 1000 different realizations with the same parameters.}
 \label{fig:ew_p}
\end{figure}

In the paramagnetic phase ($J_0< J_{0c}$), the initial values $p(0), q(0)$ 
have only a small effect on the number of defaults and a small difference between them leads to negligible consequences as shown in Figure \ref{fig:ew_p}. This figure 
presents the evolution of $ND(t)$ for four different starting points $(p(0), q(0))$ in the paramagnetic phase
with zero panic field ($H=0)$. In both cases, the number of defaults grows linearly in time with similar rate.

\begin{figure}
 \centering
 \includegraphics[scale=0.4, angle=-90]{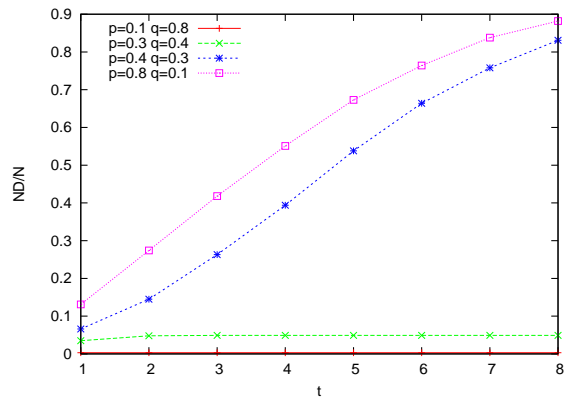}
 \caption{Same as Figure \protect\ref{fig:ew_p} for $J_0=0.005$, $\sigma_J=0.001$ corresponding to the ``ferromagnetic'' phase in which collective effects are enhanced.}
 \label{fig:ew_f}
\end{figure}

In contrast, for $J_0 \geq J_{0c}$ corresponding to the ferromagnetic phase with $H=0$, initial conditions
play a strong role in determining the subsequence evolution. Fig. \ref{fig:ew_f} illustrates
the typical situation in which two very different scenarios develop for the cumulative number of
defaults for systems starting with different initial conditions. The upper curves
in Fig. \ref{fig:ew_f} favors $s=-1$. This slight breaking of symmetry in the initial probability condition
 is sufficient to accelerate tremendously the rate of default, compared to the situation shown in Figure \ref{fig:ew_p}.
In the case $p(0)=0.3, q(0)=0.4$ corresponding to a small bias towards the ECRG change $s=+1$,
the cooperative ferromagnetic phase has the effect of actually quenching the crisis, i.e., making it
less severe with a rapid saturation to a total number of defaults more than half that obtained
for the same initial conditions in the paramagnetic phase shown in Figure \ref{fig:ew_p}.

\begin{figure}
 \centering
 \includegraphics[scale=0.4, angle=-90]{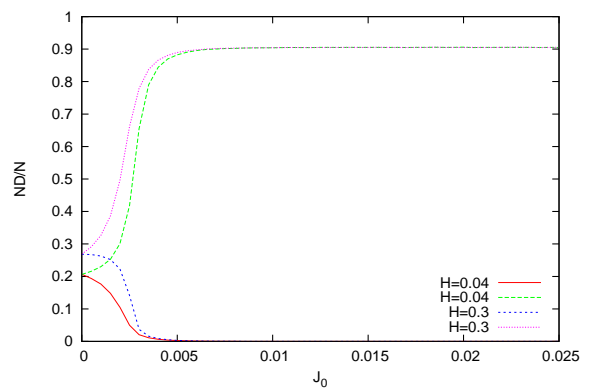}
 \caption{Cumulative number $ND$ of defaults as a function of the average coupling strength $J_0$ for pairs of different initial conditions ($p=0.1$, $q=0.1$) and ($p=0.6$, $q=0.1$) and two values of the panic field indicated in the figure ($H=0.04$ and $H=0.3$).}
 \label{fig:atr1}
\end{figure}

\begin{figure}
 \centering
 \includegraphics[scale=0.4, angle=-90]{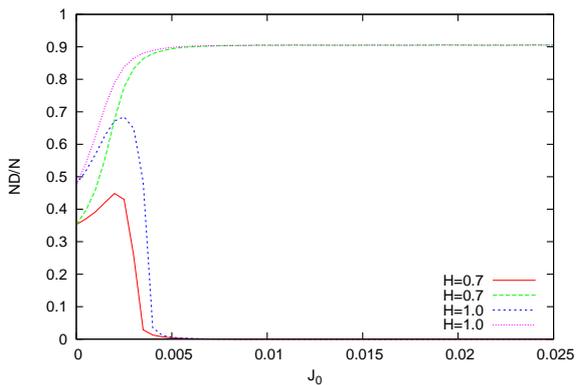}
 \caption{Same as Figure \protect\ref{fig:atr1} except for the two values of the panic field here equal to 
 $H=0.7$ and $H=1.0$.}
 \label{fig:atr2}
\end{figure}

Figures \ref{fig:atr1} and \ref{fig:atr2} illustrate further the drastically increased sensitivity to different initial 
conditions, as the average coupling strength $J_0$ crosses the critical region $J \simeq J_{0c}$, in the presence
of a non-zero panic field $H$. As shown in these Figures, different starting points may lead to opposite default scenarios, seen as a pair of two branches. As the panic field increases to rather large values, 
an interesting effect is visualized in Fig. \ref{fig:atr2}: in the absence of bias in the initial conditions ($p(0)=q(0)=0.1$),
the number of defaults is maximum close to the critical point $J_{0c}$ while going to zero for large $J_0$'s.
In contrast, when a bias is present ($p(0)=0.6, q(0)=0.1$) in favor of $s=-1$, the number of defaults 
increases rapidly to an almost complete collapse of the economy as the coupling strength increases.

\subsection{Direct impact of the phenomenon of  ``spontaneous symmetry breaking''}

\begin{figure}
 \centering
 \includegraphics[scale=0.4, angle=-90]{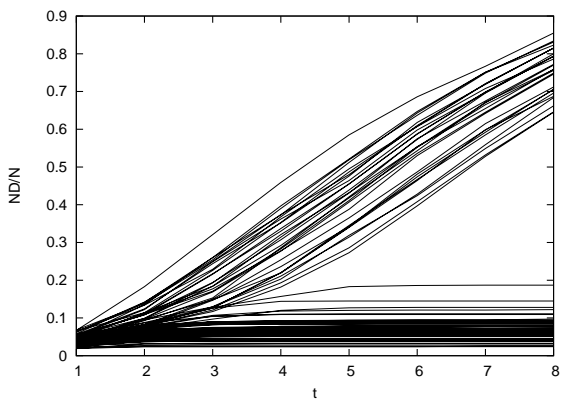}
 \caption{Time evolution of an ensemble of 100 realizations of an economy with the same parameters $J_0=0.006$ with symmetric initial conditions $p(0)=q(0)=1/3$.}
 \label{fig_ssb}
\end{figure}

These effects described in the previous subsection are in essence all the 
children of the ``spontaneous symmetry breaking'' occurring for 
$J_0 > J_{0c}$ mentioned in subsection \ref{thnbw}, amplified by two additional symmetry
breaking in the initial conditions and in the presence of the panic field. 
To make the point clearer, Figure \ref{fig_ssb} shows the time evolution of an ensemble of 
100 realizations of an economy with the same parameters $J_0=0.006$, with symmetric
initial conditions $p(0)=q(0)=1/3$. We can observe that about 40\% of the economies 
bifurcate to a large number of defaults while the other 60\% exhibit a much milder crisis.
Nothing but the phenomenon of spontaneous symmetry breaking differentiates these
different realizations. The conclusion is that, for systems functioning close to and above a
critical point, a large variability can be observed with, under the same conditions, either
a very severe or a very mild crisis unfolding. The later would not doubt be hailed as
the success of better regulations or better policy or risk management, while this
huge difference is inherent to the dynamics of systems driven by collective effects.

The pattern of time evolutions shown in figure \ref{fig_ssb}  can be interpreted
from the diagrams of figure \ref{fig:mf_iter}, specifically the cases
(${\tilde J_0} =3.5$, $H=0.1$) and (${\tilde J_0} =5.0$, $H=1.0$) in which the dark regions
correspond to states ordered along the positive or the neutral axis.
In such regimes, the number of up-going trajectories
leading to a large occurrence of firm defaults is about 40\% of all trajectories. This
40\% fraction consists of a fraction of 33\% of trajectories 
associated with the negative ordered phase and of a number of
trajectories that run first in the paramagnetic phase until
they become ordered in the neutral phase. 
Since the paramagnetic phase allows stochastic movements of firm rating
to $R=0$, a number of firms are subjected to this effect.

\section{Impact of rescue policy}

Until now, our model has implemented the negative psychological effect of defaults occurring in the economy
via a ``panic'' field, which tends to enhance the downgrading of firms and the probability of default.
We have shown that the negative sentiment, modeled by the panic field $H$ which appears when the first default happens, has the effect of destabilizing the economy and of increasing significantly the number of defaults. 
As discussed in the introduction, the inspiration of our model and subsequent analysis was the
occurrence of Lehman Brothers bankruptcy and the controversial decision by the US Treasury 
and the Federal Reserve to let Lehman Brothers fail. Our model provides a simple playground
to test what would be the effect of a policy bailing out defaulting firms in the hope of regaining stability.

We thus implemented a rescue policy in our simulations by resetting the first $B$ firms that 
default when their rating reach $R=0$ with an effective credit rating grade (ECRG) $R$
randomly chosen between $1$ and $7$ and with ECRG changes $s$ randomly set between $-1, 0$ and $+1$
with the same probabilities. Because the rescue occurs before the information that $R=0$ can spread
within the network, it corresponds to a bail-out action just before an agent defaults.

\begin{figure}
 \centering
 \includegraphics[scale=0.4, angle=-90]{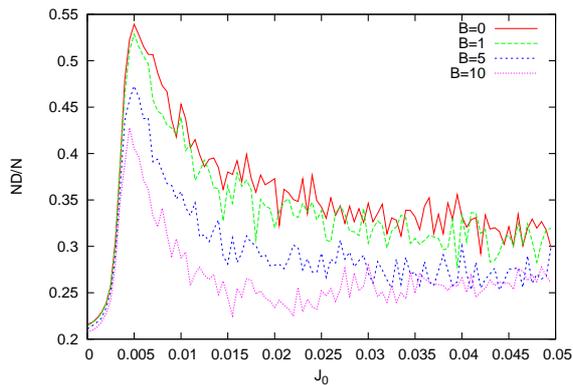}
 \caption{Cumulative number $ND(J_0)$ of defaults as a function of the average
 coupling strength $J_0$, for $H=0.08$ and different rescue parameter $B$.}
 \label{fig:rat1}
\end{figure}

\begin{figure}
 \centering
 \includegraphics[scale=0.4, angle=-90]{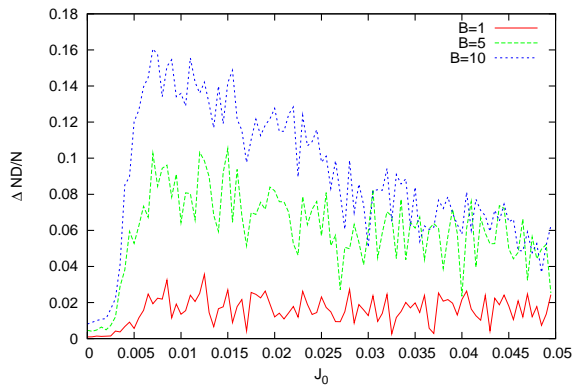}
\caption{Dependence as a function of $J_0$ of $\Delta ND /N = (ND(B=0)-ND(B))/N$, defined as the fraction
of firms that avoid bankruptcy as a consequence
of the policy to bail out the first $B$ first defaulting firms. The panic field is set to $H=0.08$
and three different rescue parameter $B$ are investigated. }
 \label{fig:rozner}
\end{figure}

\begin{figure}
 \centering
 \includegraphics[scale=0.4, angle=-90]{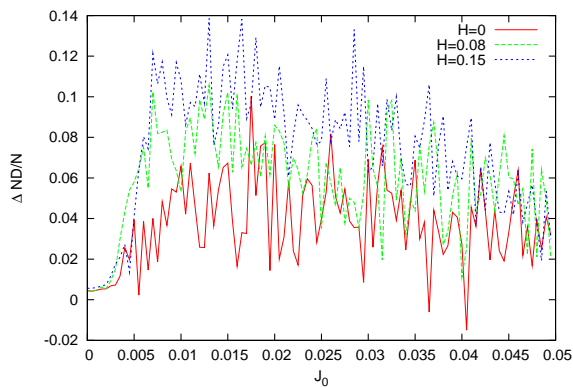}
\caption{Same as Figure \protect\ref{fig:rozner} for a fixed rescue parameter $B=5$
and three different panic fields.}
 \label{fig:rozneh}
\end{figure}

Fig. \ref{fig:rat1} shows that the bailing out policy can significantly reduce
the scale of collective bankruptcies, however it cannot prevent them.  Again, the impact
of the rescue policy is strongest close to the critical point where is the economy is the
most susceptible to shocks, either negative or positive.
The impact of the rescue policy is quantified in
Figures \ref{fig:rozner}  and \ref{fig:rozneh}, which presents 
the dependence as a function of $J_0$ of the fraction $\Delta ND /N = (ND(B=0)-ND(B))/N$
of firms that avoid bankruptcy as a consequence
of the policy to bail out the first $B$ first defaulting firms.
For a fixed moderate panic field $H=0.08$, Figure \ref{fig:rozneh} shows that 
just bailing out one firm reduces by about 2\% the relative number of defaulting firms.
While seemingly not much, for our economy of 1000 firms, this amounts to save 
endogenously about 20 firms by the spill-over effect of the collective interactions in the network of firms
at the cost of just bailing out one firm. Bailing out the first 10 defaulting firms lead
to a maximum salvation of 16\% of the firms in the economy (or 160 out of the 1000). While
the (gain / cost) ratio $= 16\%/ 10$) for $B=10$  is smaller than the 
(gain / cost) ratio $= 2\%/ 1$) for $B=1$, the effect remains significant.
 
Figure \ref{fig:rozneh} shows that the impact of the rescue policy is stronger 
for larger panic fields, as would be hoped for.

\section{Conclusions}

Johnson \cite{SimonJohnson} summarizes the three main events that triggered
the severe global phase of the crisis as follows:
``On the weekend of September 13-14, 2008, the U.S. government declined
to bailout Lehman. The firm subsequently failed, i.e., did not open for
business on Monday, September 15. Creditors suffered major losses, and
these had a particularly negative effect on the markets given that
through the end of the previous week the Federal Reserve had been
encouraging people to continue to do business with Lehman...
On Tuesday, September 16, the government agreed to provide an
emergency loan to the major insurance company, AIG. This loan was
structured so as to become the company's most senior debt and, in this
fashion, implied losses for AIG's previously senior creditors; the value
of their investments in this AAA bastion of capitalism dropped 40\%
overnight... By Wednesday, September 17, it was clear that the world's financial
markets - not just the US markets, but particularly US money market
funds - were in cardiac arrest. The Secretary of the Treasury
immediately approached Congress for an emergency budgetary appropriation
of \$700bn (about 5\% of GDP), to be used to buy up distressed assets and
thus relieve pressure on the financial system...''

Inspired by these events, we have developed a simple model of the ``Lehman Brothers effect''
in which the Lehman default event is quantified as having an almost immediate
effect in worsening the credit worthiness of all financial institutions in the economic network.
This effect embodies (i) the direct consequences of the Lehman default
on its creditors as well as (ii)  the psychological impact of the realization 
that the US Treasury and Federal Reserve was ready to let fail a major financial
institution so that no one felt protected anymore and (iii) the valuation
effect that the worthiness of financial derivatives
was much less than previously estimated, leading to a global realization that 
the problem was much more severe that imagined before.
We have offered a stylized description in which all properties of a given firm can 
be captured by its effective credit rating. We have specified simple dynamics
of co-evolution of the effective credit ratings of coupled firms, that show
the existence of a global phase transition around which the susceptibility
of the system is strongly enhanced. In this context, we show that bailing out
the first few defaulting firms does not solve the problem, but does alleviate
considerably the global shock, as measured by the fraction of firms that
are not defaulting as a consequence. We quantify a more than ten-fold
effect: bailing out the first (respectively the first ten) defaulting firm(s) saves 20 (respectively
160) more firms in an economy of 1000 firms. This amplification effect
is the direct consequence of the collective linkage between the firm credit ratings
in our economic network which is organized endogenously.

An idea to generalize the model would be to introduce  heterogeneity in size of the firms and in their
economic connections in order to make them more realistic. That could be done by introducing
 a different, non symmetric interaction matrix $J_{ij}$ in which one node would have a greater 
impact on its partners than the opposite. We leave this issue for a further studies.

\section*{Acknowledgements}
The authors acknowledge helpful discussions and exchanges with H. Gersbach, Y. Malevergne, M. Marsili 
and R. Woodard. All remaining errors are ours. PS and JAH were supported by European COST Action  
 MP0801 (Physics of Competition and Conflicts) and by the Polish Ministry of Science and Education,
 Grant No 578/N-COST/2009/0.  DS acknowledges financial support 
from the ETH Competence Center ``Coping with Crises in Complex 
Socio-Economic Systems" (CCSS) through ETH Research 
Grant CH1-01-08-2 and from ETH Zurich Foundation.  


\clearpage

\end{document}